\documentclass[prx,twocolumn,superscriptaddress,citeautoscript,showpacs,amsart,longbibliography]{revtex4-2}

\usepackage{blindtext}
\usepackage{graphicx}
\usepackage{bm}
\usepackage{epsfig}
\usepackage{amssymb}
\usepackage{amsfonts}
\usepackage{braket}
\usepackage{color}
\usepackage{epstopdf}
\usepackage{hyperref}
\usepackage{float}
\restylefloat{table}
\usepackage{bibentry}
\usepackage{color}
\usepackage{multirow}
\usepackage[caption=false]{subfig}
\usepackage{amsfonts}
\usepackage{amsthm}
\usepackage{amsmath}
\usepackage{graphicx}
\usepackage{multirow}
\usepackage{color}
\usepackage{bbold}
\usepackage{bm}
\usepackage{times}
\usepackage{amsmath,bm,amsfonts}
\usepackage{dcolumn}
\usepackage{graphicx}
\usepackage{latexsym}

\newcommand{\bpm}{\begin{pmatrix}}
\newcommand{\epm}{\end{pmatrix}}
\newcommand{\ba}{\begin{eqnarray}}
\newcommand{\ea}{\end{eqnarray}}
\newcommand{\bd}{\begin{displaymath}}

\graphicspath{{figures/}}

\begin{document}
%

\title{Observation of spin-dependent dual ferromagnetism in perovskite ruthenates}

\author{Sungsoo Hahn}
\thanks {These authors contributed equally to this work.}
\affiliation{Center for Correlated Electron Systems, Institute for Basic Science, Seoul 08826, Korea}
\affiliation{Department of Physics and Astronomy, Seoul National University, Seoul 08826, Korea}

\author{Byungmin Sohn}
\thanks {These authors contributed equally to this work.}
\affiliation{Center for Correlated Electron Systems, Institute for Basic Science, Seoul 08826, Korea}
\affiliation{Department of Physics and Astronomy, Seoul National University, Seoul 08826, Korea}

\author{Minjae Kim}
\email[Electronic address:$~$]{garix.minjae.kim@gmail.com}
\affiliation{Department of Physics, Pohang University of Science and Technology (POSTECH), Pohang, Korea}
\affiliation{Department of Chemistry, Pohang University of Science and Technology (POSTECH), Pohang, Korea}

\author{Jeong Rae Kim}
\affiliation{Center for Correlated Electron Systems, Institute for Basic Science, Seoul 08826, Korea}
\affiliation{Department of Physics and Astronomy, Seoul National University, Seoul 08826, Korea}

\author{Soonsang Huh}
\affiliation{Center for Correlated Electron Systems, Institute for Basic Science, Seoul 08826, Korea}
\affiliation{Department of Physics and Astronomy, Seoul National University, Seoul 08826, Korea}

\author{Younsik Kim}
\affiliation{Center for Correlated Electron Systems, Institute for Basic Science, Seoul 08826, Korea}
\affiliation{Department of Physics and Astronomy, Seoul National University, Seoul 08826, Korea}

\author{Wonshik Kyung}
\affiliation{Center for Correlated Electron Systems, Institute for Basic Science, Seoul 08826, Korea}
\affiliation{Department of Physics and Astronomy, Seoul National University, Seoul 08826, Korea}

\author{Minsoo Kim}
\affiliation{Center for Correlated Electron Systems, Institute for Basic Science, Seoul 08826, Korea}
\affiliation{Department of Physics and Astronomy, Seoul National University, Seoul 08826, Korea}

\author{Donghan Kim}
\affiliation{Center for Correlated Electron Systems, Institute for Basic Science, Seoul 08826, Korea}
\affiliation{Department of Physics and Astronomy, Seoul National University, Seoul 08826, Korea}

\author{Youngdo Kim}
\affiliation{Center for Correlated Electron Systems, Institute for Basic Science, Seoul 08826, Korea}
\affiliation{Department of Physics and Astronomy, Seoul National University, Seoul 08826, Korea}

\author{Tae Won Noh}
\affiliation{Center for Correlated Electron Systems, Institute for Basic Science, Seoul 08826, Korea}
\affiliation{Department of Physics and Astronomy, Seoul National University, Seoul 08826, Korea}

\author{Ji Hoon Shim}
\affiliation{Department of Physics, Pohang University of Science and Technology (POSTECH), Pohang, Korea}
\affiliation{Department of Chemistry, Pohang University of Science and Technology (POSTECH), Pohang, Korea}

\author{Changyoung Kim}
\email[Electronic address:$~$]{changyoung@snu.ac.kr}
\affiliation{Center for Correlated Electron Systems, Institute for Basic Science, Seoul 08826, Korea}
\affiliation{Department of Physics and Astronomy, Seoul National University, Seoul 08826, Korea}

\date{\today}

\begin{abstract}

We performed {\it{in-situ}} angle-resolved photoemission spectroscopy (ARPES) and spin-resolved ARPES (SARPES) experiments to investigate the relationship between electronic band structures and ferromagnetism in SrRuO$_3$ (SRO) thin films. Our high quality ARPES and SARPES results show clear spin-lifted band structures. The spin polarization is strongly dependent on momentum around the Fermi level, whereas it becomes less dependent at high-binding energies. This experimental observation matches our dynamical mean-field theory (DMFT) results very well. As temperature increases from low to the Curie temperature, spin-splitting gap decreases and band dispersions become incoherent. Based on the ARPES study and theoretical calculation results, we found that SRO possesses spin-dependent electron correlations in which majority and minority spins are localized and itinerant, respectively. Our finding explains how ferromagnetism and electronic structure are connected, which has been under debate for decades in SRO.

\end{abstract}
\maketitle

The study on ferromagnetism has been at the core of solid state physics. For example, the nature and origin of ferromagnetism in various materials have been debated for decades. While some are explained within the picture of local magnetism, others are believed to have an itinerant nature~\cite{vannette2008distinguishing,goodenough1967narrow,kubler2017theory,stoner1938collective}. A fundamental difficulty in its interpretation is that magnetism requires both local and itinerant characters; perfectly itinerant electrons may not have ferromagnetism and local magnetism needs a certain degree of delocalization for the exchange interaction~\cite{dordevic2001hybridization,fulde1995electron}. Electronic structure or, more specifically, spin-polarized single-particle spectral functions can provide vital information on the issue. For example, observation of dispersive spin-splitting bands should be a strong evidence for itinerant magnetism while momentum-independent broad spectral function suggests local magnetism~\cite{xu2020signature, schmitt2012itinerant}. However, results of spin-polarized electronic structure studies are hard to be found.

\begin{figure*}[]
	\includegraphics[width=\linewidth]{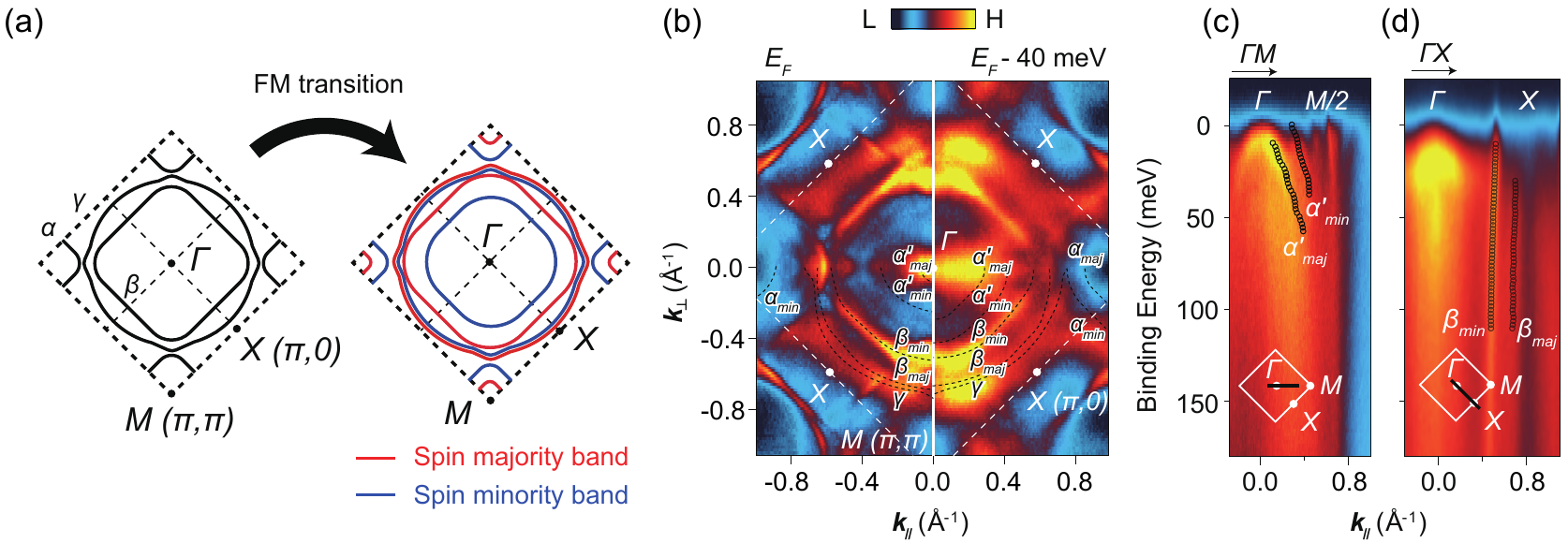}
	\centering
	\caption{Electronic structures of SrRuO$_3$ (SRO) thin films. (a) Schematic Fermi surfaces of SRO thin films without folded bands. Under finite magnetization (M),  $\alpha$, $\beta$, and $\gamma$ bands split into majority- (red) and minority-spin (blue) bands. (b) Constant energy maps at the Fermi level ($E_F$) (left) and binding energy of $40$~meV (right) are measured by angle-resolved photoemission spectroscopy (ARPES) at 10~K. Each map is obtained by integrating over an energy window of $\pm$~$4$~meV around the corresponding energy. Pseudocubic Brillouin zone is marked with white dashed lines. Majority- and minority-spin $\alpha$ ($\alpha_{maj}$ and $\alpha_{min}$), folded majority- and minority-spin $\alpha^\prime$ ($\alpha^\prime_{maj}$ and $\alpha^\prime_{min}$), majority- and minority-spin $\beta$ ($\beta_{maj}$ and $\beta_{min}$), and $\gamma$ bands are shown with black dashed lines. (c-d) High-symmetry cuts along (c) $\Gamma$-${\it M}$ and (d) $\Gamma$-${\it X}$ lines. Lorentzian functions are used to fit the peak positions as marked with black circles. Each cut is denoised with a deep learning-based statistical method~\cite{kim2021deep}. The insets show the cut direction in the Brillouin zone.}
	\label{fig:1}
\end{figure*}

In the study of ferromagnetism in terms of electronic structures, a perovskite strontium ruthenate SrRuO$_3$ (SRO) can be a good candidate material. It is a proto-typical ferromagnet and is frequently used as ferromagnetic (FM) electrodes~\cite{koster2012structure,eom1993fabrication} for transition metal oxide electronic devices~\cite{junquera2003critical}. Its ferromagnetism is believed to stem from itinerant electrons~\cite{singh1996electronic}, which has been utilized in the interpretation of the anomalous Hall effect~\cite{haham2011scaling,sohn2021stable} and optical conductivity \cite{jeong2013temperature}. In fact, the non-integer magnetic moment of the Ru ion of 1.5~$\mu_B$~\cite{longo1968magnetic,dodge1999temperature}, smaller than expected in the localized electron picture, appears to secure the itinerant magnetism. On the other hand, the large on-site Coulomb interaction of U~$\sim$~$2$~eV~\cite{fujioka1997electronic,takizawa2005manifestation,toyota2005thickness} suggests that there should be a significant portion of local magnetism as well in this material. This issue can be investigated through electronic structure studies as mentioned above.

\begin{figure*}[]
\textsc{}	\includegraphics[width=\linewidth]{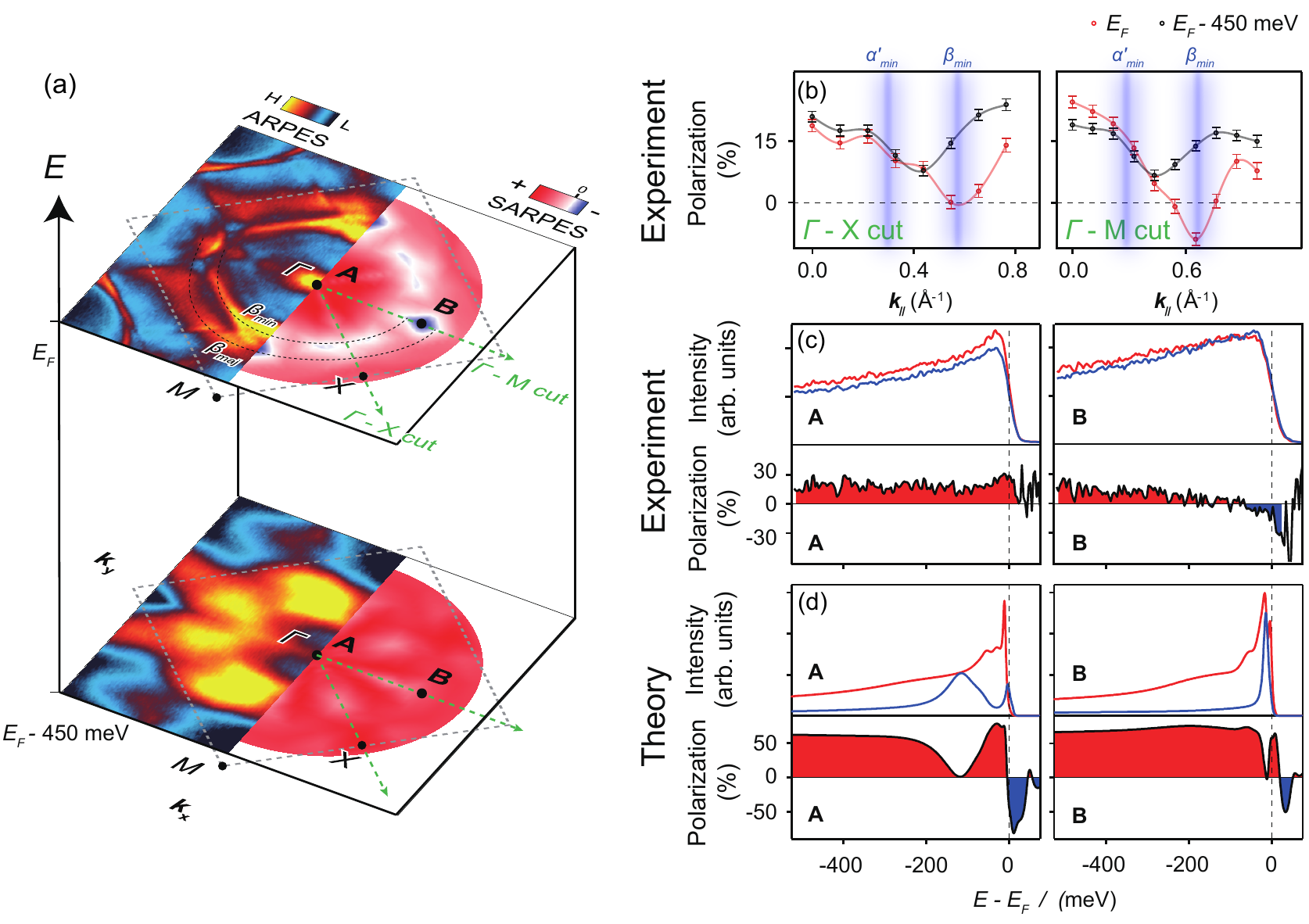}
	\vspace{-0.5cm}
	\caption{Spin-resolved electronic structures of SRO thin films. SARPES measurements were conducted at 20~K. (a) Spin-integrated and resolved constant energy maps at $E_F$ and $E_F$~-~$450$~meV measured by ARPES and SARPES. The pseudocubic Brillouin zone is indicated by gray dashed lines. Majority- and minority-spin $\beta$ ($\beta_{maj}$ and $\beta_{min}$) bands are marked with black dashed lines. Both spin polarizations are observed near $E_F$, whereas only a single spin polarization is observed in the high binding energy (HBE) region. (b) Momentum-dependent spin polarization along the high-symmetry lines ($\Gamma$-${\it X}$ and $\Gamma$-${\it M}$ as marked with green dashed lines in (a)). Red and black circles represent the spin polarizations at $E_F$ and $E_F$~-~$450$~meV, respectively. Integration windows of $E_F$~-~$25$~meV~$\pm$~25~meV and $E_F$~-~$450$~meV$\pm$~50~meV are used to obtain spin polarizations near $E_F$ and in the HBE region, respectively. The positions of $\alpha^\prime_{min}$ and $\beta_{min}$ bands are marked with blue shaded lines. (c) Experimental and (d) theoretical results of spin-resolved energy distribution curves at the $\Gamma$ point (left) and (-~$5\pi$/8, $5\pi$/8) ($\bf{A}$ and $\bf{B}$ points, respectively, in (a)) (right). Plotted in the bottom half of each panel is the spin polarization, $I$~=~$(I_{maj}-I_{min})/(I_{maj}+I_{min})$, where $I_{maj}~(I_{min})$ is the intensity of spin-polarized photoelectrons for the up- (down-) spin electrons obtained at each energy step.}
	\vspace{-0.5cm}
	\label{fig:2}
\end{figure*}

In principle, the issue may be addressed with spin-dependent electronic structure information from spin- and angle-resolved photoemission spectroscopy (SARPES) measurements~\cite{fujiwara2018origins}. Verification of such characteristics in a spin- and momentum-resolved way would be an important step towards understanding the ferromagnetism in SRO as well as other materials~\cite{petrovykh1998spin}. In this Letter, we report the results of {\it in-situ} angle-resolved photoemission spectroscopy (ARPES) and SARPES studies of epitaxial SRO thin films. 15~unit-cell (uc) SRO thin films grown on (LaAlO$_3$)$_{0.3}$(Sr$_2$TaAlO$_6$)$_{0.7}$ (LSAT) are found to have a moderate coercive field strength~\cite{xia2009critical}, allowing us to do ${\it in}$-${\it situ}$ magnetization with a permanent magnet. Our ARPES measurements show clear spin-split band structures. Spin polarization of the spin-split bands are obtained with SARPES and compared with the result of dynamical mean-field theory (DMFT) calculation. We found that both localized and itinerant electrons contribute to ferromagnetism, exhibiting spin-dependent electronic couplings which lead to itinerant and localized electrons for minority and majority spins, respectively. 

\begin{figure}[]
\includegraphics[width=\linewidth]{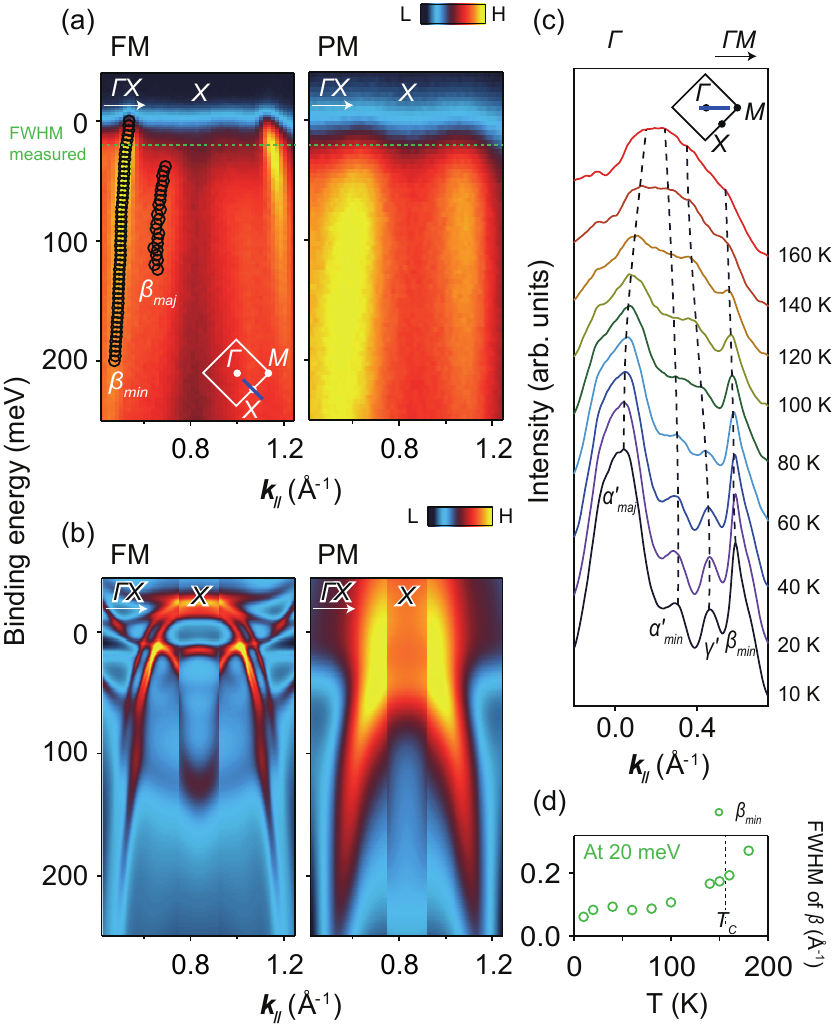}
\centering
\caption{Temperature-dependent spin-polarized band dispersions. (a) The $\Gamma$-${\it X}$ cut (left) below 10~K and (right) above {\it T$_C$} (160~K). Each cut is denoised with a deep learning-based statistical method~\cite{kim2021deep}. (b) Calculated ARPES spectra of (left) ferromagnetic (FM) and (right) paramagnetic (PM) states based on dynamical mean-field theory (DMFT) calculation along the $\Gamma$-${\it X}$ line. (c) Momentum distribution curves (MDCs) at $E_F$ along the $\Gamma$-${\it M}$ line for various temperatures between 10 and 160~K. Black dashed lines are guides to the eye for the temperature-dependent behavior of $\alpha^\prime$, $\beta$, and $\gamma^\prime$ bands. (d) Temperature dependence of the full width at half maximum (FWHM) of $\beta_{min}$ band peaks at $E_F$~-~$20$~meV. The values are obtained from the Lorentzian fit of MDC peaks.}
\label{fig:3}
\end{figure}

Details of experimental methods and DMFT+DFT calculation~\cite{georges1996dynamical, kotliar2006electronic} are described in the Supplementary Materials (SM)~\cite{SM}. Figure 1(a) schematically shows the Fermi surface topology of SRO in which three $t_{2g}$ bands ($\alpha$, $\beta$, and $\gamma$) cross the Fermi Level ($E_F$)~\cite{sohn2019sign, shai2013quasiparticle}. The bands in the non-magnetic state (left) spin split into six bands as in the right figure of Fig. 1(a). Although SRO is FM and the band splitting is expected to exist, the spin-split bands have not yet been experimentally observed in SRO~\cite{shai2013quasiparticle}. With ${\it in}$-${\it situ}$ ARPES measurements on a high-quality thin film, we carefully examined the fermiology and successfully observed spin-split band structures, {\it i.e.} majority- and minority-spin bands. 

Shown in Fig. 1(b) are constant energy maps of SRO thin films at $E_F$ (left half) and at $E_F$~-~$40$~meV (right half). The bands are named in accordance with those in previous reports~\cite{shai2013quasiparticle,sohn2019sign}. $\alpha^\prime$ (folded $\alpha$) and $\beta$ bands split into majority- and minority-spin bands as shown in Fig. 1(b). Hole-like majority- and minority-spin bands of $\alpha^\prime$ ($\alpha^\prime_{maj}$ and $\alpha^\prime_{min}$, respectively) are observed around the $\Gamma$ point while electron-like majority- and minority-spin $\beta$ bands ($\beta_{maj}$ and $\beta_{min}$, respectively) are located outside of $\alpha^\prime$ Fermi surfaces. Note that oxygen octahedra in the real systems are distorted, which induces $\sqrt{2}$$\times$$\sqrt{2}$ or 2$\times$2 enlarged unit cells~\cite{roh2021structural,chang2011thickness,vailionis2008room}. As a result, folded bands are observed in ARPES data in addition to the bands in Fig. 1(a). The folded bands are marked with a prime symbol in Fig. 1(b) (See SM~\cite{SM} for a schematic Fermi surface with folded bands).

Figures 1(c) and (d) show high-symmetry cuts along the $\Gamma$-$M$ and $\Gamma$-$X$ directions as marked in the insets. Each cut clearly shows the majority- and minority-spin $\alpha^\prime$ and $\beta$ bands. In order to obtain the band dispersions, we fitted momentum distribution curves (MDCs) with Lorentzian functions and determined the MDC peak positions as shown with black circles in the figures. We speculate that $\alpha^\prime$ ($\beta$) band is better observed in the $\Gamma$-$M$ ($\Gamma$-$X$) cut due to the symmetry-related matrix element effects~\cite{iwasawa2010interplay, iwasawa2012high}.

\begin{figure}[]
\includegraphics[width=\linewidth]{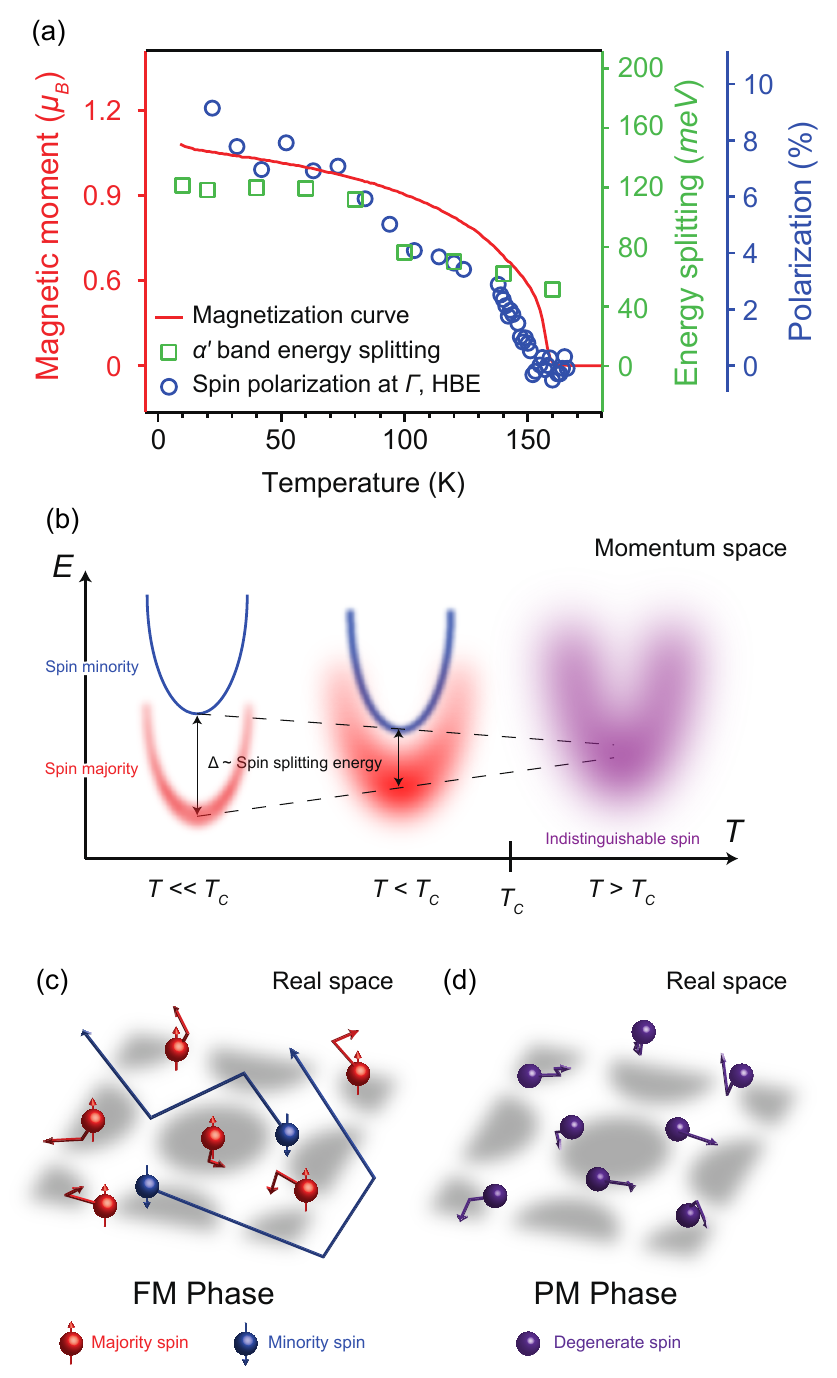}
\caption{Spin-dependent itinerant and localized ferromagnetism. (a) Temperature-dependent magnetic moment (red solid line), energy splitting between spin-polarized bands (green squares), and spin polarization near $E_F$~-~$450$~meV (HBE) at the $\Gamma$ point (blue circles), respectively. Energy splitting is obtained from the difference between $\alpha^\prime_{maj}$ and $\alpha^\prime_{min}$ in Fig. 3(c). (b) A schematic description for temperature-dependent spin-polarized bands. Black dashed lines connect the band bottom of each band. A red band (blue band) represents the majority- (minority-) spin band, and a purple band represents indistinguishable spin bands. (c, d) A schematic description for movements of electrons in real space in FM and PM phases. The majority (minority) spins are localized (itinerant) in the FM phase, whereas spin-degenerate electrons are localized.
}
\label{fig:4}
\end{figure}

To confirm the spin-polarized nature of the bands, we performed ${\it in}$-${\it situ}$ SARPES on the SRO thin film. Figure 2(a) shows spin-dependent constant energy maps at $E_F$ and $E_F$~-~$450$~meV with the corresponding constant energy map measured with ARPES. At $E_F$, positive spin polarization (red) is observed near the $\Gamma$ point, but it turns negative (blue) away from the $\Gamma$ point. As Fig. 2(b) shows, we found that the $\alpha^\prime_{min}$ and $\beta_{min}$ bands are negatively spin-polarized. The spin polarization turns positive again when the momentum increases beyond the two bands. Surprisingly, the momentum dependence of spin polarization almost vanishes at high-binding energy (HBE) of $E_F$~-~$450$~meV; the spin polarization is entirely positive regardless of the electron momentum.

The energy-dependent behavior of the spin polarization can be better seen in spin-dependent energy distribution curves and spin polarization $I$~=~$(I_{maj}-I_{min})/(I_{maj}+I_{min})$ from $\Gamma$ ({\bf A}) and (-~$5\pi$/8, $5\pi$/8) ({\bf B}) plotted in Fig. 2(c). At {\bf B}, a negative spin polarization is observed near $E_F$ which is due to the contribution from $\alpha^\prime_{min}$ and $\beta_{min}$ bands. On the other hand, the near $E_F$ spin polarization is positive at {\bf A} due to the contribution from the $\alpha^\prime_{maj}$ band. In the HBE region, the spin polarization remains positive regardless of the momentum which is amassed to form majority spins. This behavior is consistent with what is observed in Fig. 2(a). The experimental results are corroborated with DMFT calculation results in Fig. 2(d); the spin polarization changes its sign near $E_F$, whereas it is constantly positive at HBEs. Based on the present and previous DMFT calculation results~\cite{kim2015nature}, we assert that the momentum independent positive spin polarization at HBEs stems from the localized character of the majority-spin band. A localized electron band is expected to be broad and weakly dispersive~\cite{damascelli2003angle,kwon2019lifshitz}. Therefore, the majority-spin band is also expected to provide momentum-independent spin polarization. The broader and incoherent $\beta_{maj}$ band in Fig. 1(d) indeed indicates its localized character, different from the $\beta_{min}$ band.

The spin-split bands are expected to have strong temperature-dependent behaviors~\cite{pickel2010magnetic,fujiwara2018origins}, especially approaching the Curie temperature ($T_C$). Figure 3(a) shows the high-symmetry cuts along the $\Gamma$-$X$ direction at 10 and 160~K, below and above $T_C$ (See SM~\cite{SM} for the $\Gamma$-$M$ cut). Both $\beta_{maj}$ and $\beta_{min}$ are observed at 10~K, whereas it is difficult to distinguish the two bands at 160~K. Overall, the temperature-dependent behavior is that the spin-split gap decreases, and bands become broader as the temperature increases. Eventually, the spin-split bands become indistinguishable near $T_C$. Figure 3(b) shows band dispersions of bulk SRO along the $\Gamma$-$X$ high-symmetry cut from DMFT calculation. $\beta_{maj}$ and $\beta_{min}$ bands are seen in the FM state while the two bands are degenerate in the paramagnetic (PM) state.

In order to investigate the temperature dependence more closely, we plot in Fig. 3(c) $\Gamma$-$M$ MDCs at $E_F$ for various temperatures. The momentum space distance between $\alpha^\prime_{maj}$ and $\alpha^\prime_{min}$ peaks becomes smaller as the temperature increases. Eventually, the two $\alpha^\prime$ peaks merge each other and become indistinguishable near $T_C$. Moreover, the peaks tend to be broader and incoherent with increasing temperature, as shown in the temperature-dependent full width at half maximum (FWHM) plot of $\beta_{min}$ in Fig. 3(d). 

We may compare the temperature-dependent SAPRES results with the magnetization of films. Plotted in Fig. 4(a) is the magnetic moment per Ru ion (from SQUID measurements, solid red line), $\alpha^\prime$ band splitting (green squares) and spin polarization at the $\Gamma$ point (blue circles) (See SM~\cite{SM} for details on the band splitting energy). The $T_C$ is 156~K, which is obtained from the SQUID data. It is shown that the spin polarization follows the magnetic moment curve, vanishing near the $T_C$. The $\alpha^\prime$ band splitting has a temperature dependence similar to that of the magnetic moment in the low-temperature region, but cannot be precisely determined near $T_C$ as the two bands cannot be discerned.

Cartoonistic summary of the observed temperature-dependent behavior of the spin-resolved electronic structure is given in Fig. 4(b). Overall, the majority-spin band is more incoherent than the minority-spin band. In the low-temperature region, the majority- and minority-spin bands have a large spin-split gap, $\Delta$. As the temperature increases, the bands become incoherent and broader with decreasing $\Delta$. Above $T_C$, the two spin bands become more incoherent and cannot be distinguished. At present, it is not clear if the two bands immediately merge at $T_C$ or remain split above $T_C$~\cite{jeong2013temperature}. The temperature-dependent behavior in the real space is schematically shown for the FM (below $T_C$) and PM (above $T_C$) phases.

An interesting aspect of the temperature dependence is that the majority-spin band is more incoherent than the minority-spin band, supported both experimentally and theoretically~\cite{kim2015nature}. In general, bands become incoherent as the electronic correlation gets stronger~\cite{damascelli2003angle,kwon2019lifshitz}. With self-energy analysis in our DMFT calculation (See SM~\cite{SM} for details), we found that majority-spin bands are more incoherent and correlated than minority-spin bands. Thus, majority-spin electrons have shorter mean-free path, $\it{l}$($\omega$), in comparison to minority-spin electrons, which follows $l(\omega) = v^{\ast}(\omega) \times \tau(\omega)$, where $v^{\ast}(\omega)$ and $\tau(\omega)$ = (-$\textsl{Z}$~Im$\Sigma(\omega))^{-1}$ are renormalized quasiparticle velocity and quasiparticle lifetime, respectively, as depicted in Fig. 4(c). Meanwhile, in the PM state, the majority- and minority-spin bands merge, and the band structure becomes very incoherent. Therefore, we expect that the electron mean-free path in the PM state is sufficiently short as schematically shown in Fig. 4(d). This view is also supported by DMFT calculation results (See SM~\cite{SM} for details).

In summary, we have investigated the connection between the ferromagnetism and electronic structure in strontium ruthenate thin films. By combining ARPES and SARPES studies, we report the clear band structure and its temperature- and spin-dependent behaviors, which show a good agreement with DMFT calculation results. In particular, we demonstrate that majority- and minority-spin electrons are acting differently and induce the dual ferromagnetism having (spin-dependent) locality and itinerancy at the same time. A decades-long problem in the magnetism in SRO may be explained within our picture, which sheds light on the connection between the electronic structure and magnetism. Our work proposes a new way to explore the nature of ferromagnetism in correlated materials.

We thank Y. Ishida for fruitful discussion. This work is supported by the Institute for Basic science in Korea (Grant No. IBS-R009-D1, IBS-R009-G2). The work at Postech was supported by NRF (Grant No. NRF-2021R1A2C2010972)


\end{document}